\documentclass[onecolumn,nofootinbib]{revtex4}
\usepackage{setspace}
\usepackage{amsmath,amssymb}
\usepackage{graphicx}
\usepackage{dcolumn}
\usepackage{bm,color}
\usepackage{hyperref}
\usepackage{accents}
\usepackage{amssymb,float}
\usepackage{amsmath}
\usepackage{multirow}
\usepackage{tikz}
\usepackage{subcaption}
\usepackage[rightcaption]{sidecap}
\sidecaptionvpos{figure}{t}
\usepackage{enumerate}

\newcolumntype{P}[1]{>{\centering\arraybackslash}p{#1}}
\newcolumntype{M}[1]{>{\centering\arraybackslash}m{#1}}
\hypersetup{
    colorlinks=true,
    linkcolor=red,
    filecolor=magenta,      
    citecolor=blue
}

\newcommand{\dd}{\mathrm{d}}
\begin{document}

\title{Dynamical systems in Einstein Gauss-Bonnet gravity}

\author{Konstantinos F. Dialektopoulos}
\email{kdialekt@gmail.com}
\affiliation{Department of Physics, Nazarbayev University, 53 Kabanbay Batyr avenue, 010000 Astana, Kazakhstan}
\affiliation{Laboratory of Physics, Faculty of Engineering, Aristotle University of Thessaloniki, 54124 Thessaloniki, Greece}

\author{Jackson Levi Said}
\email{jackson.said@um.edu.mt}
\affiliation{Institute of Space Sciences and Astronomy, University of Malta, Malta, MSD 2080}
\affiliation{Department of Physics, University of Malta, Malta}

\author{Zinovia Oikonomopoulou}
\email{zhnobia.oikonomopoulou.21@um.edu.mt}
\affiliation{Institute of Space Sciences and Astronomy, University of Malta, Malta, MSD 2080}

\date{\today}

\begin{abstract}
In this work we explore the dynamical system phase space of Einstein-Gauss-Bonnet theory in the cosmological minisuperspace. This approach binds the main features of the theory through a system of autonomous differential equations, in the context of a flat Friedmann–Lema\^{i}tre–Robertson–Walker spacetime. We analyze the critical points that feature in this system to assess their stability criteria. The phase space of this form of scalar-tensor gravity is very rich due to the fourth-order contributions of the Gauss-Bonnet invariant together with the second order contribution of the scalar field together with their coupling dynamics. We find additional critical points as compared with previous works in the literature which may be important for understanding the larger evolution of standard background cosmology within this class of gravitational models.
\end{abstract}

\maketitle

\section{Introduction}\label{sec:intro}

Over the last decades, general relativity has been extremely successful in explaining cosmological observations, through its reformulation in the standard cosmological model ($\Lambda$CDM) \cite{misner1973gravitation,Clifton:2011jh,Aghanim:2018eyx}. Here, an initial big bang undergoes complex early Universe processes to eventually give a late time Universe that is undergoing accelerated expansion driven by dark energy described by a cosmological constant \cite{Riess:1998cb,Perlmutter:1998np}. Despite these successes, $\Lambda$CDM continues to express fundamental problems due to the nature of the cosmological constant such as fine-tuning, among many others \cite{RevModPhys.61.1,Appleby:2018yci,Ishak:2018his}. The other core ingredient to $\Lambda$CDM is that of cold dark matter (CDM) which dominates on galactic scales, and defines the large scale structure of the Universe. Similarly, despite its theoretical success, CDM remains elusive to direct measurements in particle detectors \cite{Baudis:2016qwx,Bertone:2004pz}. On the observational side of the cosmological model building, $\Lambda$CDM faces a new challenge in the form of cosmological tensions wherein the predictions of the standard model using early time data from the Planck \cite{Aghanim:2018eyx} and ACT \cite{ACT:2020gnv} observatories are in tension with some local measurements of the Hubble constant \cite{Bernal:2016gxb,DiValentino:2020zio,DiValentino:2021izs,Riess:2019cxk,Wong:2019kwg}. The problem appears to be growing and may also pollute other measurements such as that of the evolution of large scale structure \cite{Abdalla:2022yfr,DiValentino:2020vvd}.

An alternative scenario is the quintessence model which incorporates a minimally coupled, canonical scalar field responsible for the current accelerated expansion \cite{Caldwell:1997ii}. Quintessence models are advantageous for a number of scenarios. One of these is the tracker nature of the quintessence field which gives the scalar field a behavior that mimics dark energy once the matter-radiation equality point in time is passed \cite{Tsujikawa:2013fta}. This makes the model almost indistinguishable from $\Lambda$CDM at late times but dynamical at early times. In this way, the early time physics can be modified to produce models that are more amenable to being made to agree with observational measurements. Despite the fact that the standard quintessence is indeed a plausible model, still it does not incorporate a phantom regime, which is something that may be advantageous to explaining the latest indications by the Planck data releases \cite{Aghanim:2018eyx}.

Another perspective already present in the literature, is the modification of GR in various ways in order to describe the evolution of the Universe. Among the plethora of those models, there is a specific case that one can actually modify gravity by adding quadratic order terms of the curvature tensor in the Lagrangian. This special class of modified gravity falls under the Lovelock's Lagrangian formulation in $n$-dimensions \cite{Clifton:2011jh}. In this scheme, there exists a particular quadratic combination of the curvature tensor, known as the Gauss-Bonnet (GB) term which is a topological invariant in four dimensions space-time. In addition, it can preserve the second order character of the equations of motion. Even though the GB term is a total divergence in four dimensions, it could in fact contribute to the dynamics of a system provided that it is coupled to dynamically evolving scalar field. In that case we are dealing with the Einstein-Gauss-Bonnet (EGB) model, a scalar-tensor theory belonging to the Horndeski class \cite{Fernandes:2022zrq}. This has been investigated in Refs.~\cite{Koivisto:2006ai,Neupane:2006dp}, where the dynamical system was explored. However, this is a rich platform on which to formulate cosmological models, and moreover these works do not incorporate the recent constraints on the speed of gravitational waves, which can have an impact on the resulting dynamical system.

As with all modified gravity theories, the EGB model must be capable of describing at least most of the different evolution eras of the Universe. In the present article we address that issue by employing a dynamical system approach and discuss the resulting impact on that dynamical system when the theory is required to be compatible with the gravitational wave speed constraint. As we will see shortly, the autonomous feature of the dynamical system could be established from the start without making any assumptions about different cosmological eras. In this way, we explore the dynamical features of these classes of cosmological models through their critical points and the nature of these features. Through this analysis, we are then better able to probe the viability of specific models in the context of the critical points that they express, and whether they correlate with the nature of the real Universe. In the same trend and in order to emphasize more on the results of the current research, we decided to include adequate plots for each critical point. As a result, the resulting visualisation could highlight the key characteristics of each one of them.

Our study begins with Sec.~\ref{sec:EGB_foundations} where the theoretical framework of EBG theory is presented. Namely, the derivation of the field equations from the Lagrangian of the model, as well as the parameters and constraints for a flat FLRW Universe. In Sec.~\ref{sec:dyn_sys_ana} we construct the dynamical system of the EGB theory, discuss the complete phase space of this model and derive certain results regarding the behaviour of the equilibrium points. In Sec.~\ref{sec:phase_port} we demonstrate the phase portraits of the system in order to clarify the results from the previous section. Finally, in Sec.~\ref{sec:con} we present a discussion about the crucial results of the present article and the impact that might have on our understanding of the evolution of the Universe.

\section{Nonminimally Coupled Gauss-Bonnet Cosmology} \label{sec:EGB_foundations}

Let us begin by introducing the theory under considerations, i.e. Einstein Gauss-Bonnet gravity. Its action reads,
\begin{equation} \label{eq:2}
    S = \int \dd ^4 x \sqrt{-g} \left(\frac{R}{2}-\frac{1}{2} \,g ^{\mu\nu} \,\partial _{\mu} \phi \,\,\partial _{\nu} \phi -V(\phi) -f(\phi)\, \mathcal{G}\right) +S_{\rm matter}\,,
\end{equation}
where $$\mathcal{G}=R^2-4R_{\alpha\beta}\,R^{\alpha\beta}+R_{\alpha\beta\gamma\delta}\,R^{\alpha\beta\gamma\delta}\,,$$ is the Gauss-Bonnet term, that is a topological invariant in four dimensions, $\phi$ is a scalar field, non-minimally coupled to the GB term through an arbitrary function $f(\phi)$ and $V(\phi)$ is its potential. $S_{\rm matter}$ represents collectively the action of all matter fields. 

Let us consider a spatially flat Friedmann-Lema\^itre-Robertson-Walker (FLRW) spacetime with a line element that reads
\begin{equation}
    \dd s^2 = - N^2(t) \dd t^2 + a^2(t) \sum_{i=1}^{i=3} (\dd x^i)^2\,,
\end{equation}
where $a(t)$ is the scale factor, $N(t)$ the lapse function and $\dd x^i$ the Cartesian coordinates. The Ricci scalar and Gauss-Bonnet term take respectively the following form,
\begin{equation}
    R= \frac{6}{N^2} \left(2 H^2 +\dot{H} - \frac{\dot{N}}{N}H \right)\,, \quad \text{and} \quad \mathcal{G}= \frac{24 H^2}{N^4} \left(\dot{H}+H^2-\frac{\dot{N}}{N}H \right)\,,
\end{equation}
where we have introduced the Hubble function defined as $H=\dot{a}/a$ and over-dots denote derivatives with respect to the cosmic time.

Regarding the self-interaction potential of the scalar field, we choose to work with the exponential form, which is widely used in cosmology and has a solid theoretical justification in string theory \cite{Copeland:1997et,Kehagias_2004}. Therefore, we consider the potential to be a decreasing function of the scalar field
\begin{equation}
    V(\phi)=V_0 e^{-\lambda\phi}\,,
\end{equation}
with $V_0$ and $\lambda$ being two real, positive constants. In addition, as far as the coupling function is concerned, it has been seen in the literature \cite{Doneva:2022ewd,Antoniou:2022agj,Antoniou:2021zoy,Kanti:2019upz,Antoniou:2017acq} that the linear and the exponential functions are the ones that lead to scalarization, meaning that there exist black holes with non-trivial scalar hair. For this reason, we consider it to be
\begin{equation}
    f(\phi)=\frac{e^{k\phi}}{k}\,,
\end{equation}
where $k$ is a real, positive constant.

Substituting this into the action in Eq.~\eqref{eq:2}, the Lagrangian density of the gravitational sector takes the form
\begin{equation}\label{eq:6}
    \mathcal{L}= - a^3 N \,V(\phi) + \frac{3 a \dot{a}^2}{N} + \frac{a^3 \dot{\phi}^2}{2 N} + \frac{3}{N}\left(\ddot{a} - \frac{\dot{a}\dot{N}}{N}\right)\left(a^2 - \frac{8f(\phi)\dot{a}^2}{N^2}   \right)\,.
\end{equation}
After integration by parts, we get the point-like Lagrangian that reads
\begin{equation}
    \mathcal{L}= - a^3 N V(\phi) - \frac{3a\dot{a}^2}{N} + \frac{8\dot{a}^3 \dot{f}}{N^3} + \frac{a^3 \dot{\phi}^2}{2 N}\,.
\end{equation}

Using the Euler-Lagrange equations, we get for the scale factor, the lapse function and the scalar field, respectively as
\begin{gather}
    16 H \left(H^2 + \dot{H} \right) \dot{f}(\phi) -2 \dot{H} + V(\phi) - \frac{\dot{\phi}^2}{2} +  H^2 \left(8 \ddot{f}(\phi) - 3 \right) = 0 \,,\\ \label{eq:10} 
    3H^2 - V(\phi) -24 H^3 \dot{f}(\phi) - \frac{1}{2}\dot{\phi}^2 = 0 \,,\\
    \ddot{\phi} + 3 H \dot{\phi} + V'(\phi) + 24 H^2 f'(\phi) \left(H^2 + \dot{H} \right) = 0\,,\label{eq:9}
\end{gather}
where $\dot{ }$ represents derivative with respect to $t$ and $'$ derivative with respect to the argument. In the above, we have adopted the lapse gauge, meaning $N=1$.

Regarding the matter content of the Universe, we consider the energy-momentum tensor of a perfect fluid with isotropic pressure,
\begin{equation}
    T^\mu{}_\nu={\rm diag}\left(- \rho _M ,P_M,P_M,P_M\right)\,.
\end{equation}
The continuity equations for the mass-energy densities of non-relativistic matter and radiation are
\begin{align}
    &\dot{\rho}_m\,+ \,3 H {\rho}_m\,=\,0, \,\,\,\,\text{for non-relativistic matter}\,,\\
    &\dot{\rho}_r\,+\,4 H {\rho}_r\,=\,0,\quad\,\,\text{for radiation}\,,
\end{align}
where we have used the equation of state for radiation, that is $P_r = \rho_r/3$ and dust is pressureless.

\section{Dynamical System Analysis}\label{sec:dyn_sys_ana}

A dynamical system is any real or even artificial group of elements that evolves over time. As such, a dynamical system is built by differential equations associated with time derivatives. It follows, that a universal theory for studying the behavior of a dynamical system is extremely unlikely to exist. As a result, we should apply a variety of methods to the evolution rule governing the dynamical system to capture its characteristics\cite{katok1995introduction,strogatz2000nonlinear,abraham1992dynamics}.

The majority of the cosmological models are described by a system of non-linear differential equations and particularly in this case, since the system of equations cannot be easily solved we choose to use dynamical systems to study the theory in order to probe its evolutionary dynamics. An extensive review on dynamical systems in cosmology can be found in Refs.~\cite{Bahamonde:2017ize,wainwright_ellis_1997}.

In our study, we introduce the following phase space variables
\begin{equation}
    x_1=\frac{\dot{\phi}}{\sqrt{6}H},\,\,\,x_2=\frac{\sqrt{V}}{\sqrt{3}H},\,\,\,x_3=H^2 \,\frac{\partial f(\phi)}{\partial \phi},\,\,\,x_4=\frac{\sqrt{\rho_r}}{\sqrt{3}H},\,\,\,x_5=\frac{\sqrt{\rho_m}}{\sqrt{3}H}\,.
\end{equation}
The first two are related to the scalar field of the theory, namely to its kinetic term and potential energy respectively; the third variable involves the coupling of the GB invariant to the scalar field of the theory, while the last two variables have to do with the matter fields.

Further, we use these variables to express the dimensionless density parameters as well as the equation of state for the scalar field and the effective equation of state of the EGB model in a flat FLRW Universe
\begin{gather}
    \Omega_{\phi}=x_1^2+x_2^2,\,\,\, \Omega_{GB}=8 \sqrt{6} \,x_1 \,x_3,\,\,\, \Omega_r=x_4^2,\,\,
    \Omega_m=x_5^2\,,\\
    w_{\phi}=\dfrac{x_1^2\,-\,x_2^2}{x_1^2\,+\,x_2^2}\,,\,\,\,w_{\rm eff}=-1-\frac{2}{3}\frac{\dot H}{H^2}\,,
\end{gather}
which will be useful in understanding the late time value of the effective dark energy equation of state. In that way, we can recast the Friedmann constrain Eq.~\eqref{eq:10} as
\begin{equation}
    x_1^2\,+\,x_2^2\,+\,8\sqrt{6}\,x_1\,x_3\,+\,x_4^2\,+\,x_5^2\,=\,1\,,
\end{equation}
where we added the contribution of the matter fields as well.

Since we are equipped with the appropriate variables, our next step is to construct the equations for their evolution over time. In order for the EGB model to be valid at all times, a logarithmic time variable is taken, namely the number of e-foldings $N$, defined as
\begin{equation}
 N\, =\, \int_{t_i}^{t_f} H(t) \,dt\,.
\end{equation}
The derivatives of the variables with respect to the number of e-foldings $N$ $(\dd N = H \dd t)$ are
\begin{align}
    &\frac{\dd x_1}{\dd N}\,=\,x_1'\,=\,-3x_1\,+\,\sqrt{\frac{3}{2}} \,\lambda\, x_2^2\,-\,4 \,\sqrt{6} \,x_3 \,-\,(\,4\,\sqrt{6}\, x_3\,+\,x_1\,)\,\frac{\dot{H}}{H^2}\label{eq:x_1}\\[3pt]
    &\frac{\dd x_2}{\dd N}\,=\,x_2'\,=\,- \sqrt{\frac{3}{2}} \,\lambda x_1x_2\,-\,x_2\,\frac{\dot{H}}{H^2}\label{eq:x_2}\\[3pt]
    &\frac{\dd x_3}{\dd N}\,=\,x_3'\,=\,k\,\sqrt{6}\,x_1\,x_3\,+\,2 \,x_3\,\frac{\dot{H}}{H^2}\label{eq:x_3}\\
    &\frac{\dd x_4}{\dd N}\,=\,x_4'\,=\,-2\,x_4\,-\,x_4\,\frac{\dot{H}}{H^2}\label{eq:22}\\[3pt]
     &\frac{\dd x_5}{\dd N}\,=\,x_5'\,=\,-\frac{3}{2}\,x_5\,-\,x_5\,\frac{\dot{H}}{H^2}\,,\label{eq:x_5}
\end{align}

However, the system described in Eqs.~\eqref{eq:x_1}-\eqref{eq:x_5} features one open problem in the definition of the $\dot{H}/H^2$ fraction, which may incorporate important dynamics of the system. This fraction is assumed to be a constant in some works \cite{Chatzarakis:2019fbn}, whereas we endeavour to generalize this approach and explore potential impacts form this evaluation.

In order to determine this fraction, we adopt the results of Ref.~\cite{Odintsov:2019clh} and Ref.~\cite{Odintsov:2020sqy}, where it is demonstrated that the EGB model may eventually be compatible with GW170817 event. This is achieved if we consider that in EGB theory, the gravitational wave speed is 
\begin{equation} \label{eq:26}
    c_T^2\,=\,1\,-\,\frac{Q_f}{2 Q_t}\,, 
\end{equation}
with $Q_f = 8 \left(\ddot{f} - H \dot{f}\right)$ and $Q_t = 1 - 4 H \dot{f}$. This means that, since $c_T^2 = 1$ from GW170817, we need to have $Q_f = 0$. Taking the time derivative of Eq.~\eqref{eq:10} and substituting the derivative of the potential from Eq.~\eqref{eq:9}, while adding the contribution of the matter fields, we end up with,
\begin{equation}\label{eq:16}
    -2  \dot{H} -  \dot{\phi}^2 + 8 H^2 \left(\ddot{f} - H \dot{f} \right) +  16 H \dot{H} \dot{f}   - {\rho}_m-\frac{4{\rho}_r}{3} = 0 \,.
\end{equation}
Solving Eq.~\eqref{eq:16} for $\dot{H}/H^2$ we get
\begin{equation}\label{eq:27}
    \frac{\dot {H}}{H^2}\,=\,\frac{6\,x_1^2\,+\,4\,x_4^2\,+\,3\,x_5^2}{16\,\sqrt{6}\,x_1\,x_3\,-\,2}\,,
\end{equation}
which is a general solution that makes the theory compatible with the GW170817 event and no assumption was made whatsoever. Replacing this result in Eqs.~\eqref{eq:x_1}-\eqref{eq:x_5}, we obtain the final form of the autonomous dynamical system, that is
\begin{align}
    x_1' &=\,-3x_1\,+\,\sqrt{\frac{3}{2}} \,\lambda\, x_2^2\,-\,4 \,\sqrt{6} \,x_3 \,-\,(\,4\,\sqrt{6}\, x_3\,+\,x_1\,)\,\left(\,\cfrac{6\,x_1^2\,+\,4\,x_4^2\,+\,3\,x_5^2}{16\,\sqrt{6}\,x_1\,x_3\,-\,2}\,\right)\label{eq:28}\\
    x_2' &=\,- \sqrt{\frac{3}{2}} \,\lambda\, x_1x_2\,-\,x_2\,\left(\,\cfrac{6\,x_1^2\,+\,4\,x_4^2\,+\,3\,x_5^2}{16\,\sqrt{6}\,x_1\,x_3\,-\,2}\,\right)\label{eq:29}\\
    x_3' &=\,k\,\sqrt{6}\,x_1\,x_3\,+\,2 \,x_3\,\left(\,\cfrac{6\,x_1^2\,+\,4\,x_4^2\,+\,3\,x_5^2}{16\,\sqrt{6}\,x_1\,x_3\,-\,2}\,\right)\label{eq:30}\\
    x_4' &= -2\,x_4\,-\,x_4\,\left(\,\cfrac{6\,x_1^2\,+\,4\,x_4^2\,+\,3\,x_5^2}{16\,\sqrt{6}\,x_1\,x_3\,-\,2}\,\right)\label{eq:31}\\
     x_5' &=\,-\frac{3}{2}\,x_5\,-\,x_5\,\left(\,\cfrac{6\,x_1^2\,+\,4\,x_4^2\,+\,3\,x_5^2}{16\,\sqrt{6}\,x_1\,x_3\,-\,2}\,\right)\,,\label{eq:32}
\end{align}

It is evident that by employing the gravitational wave speed constraint, we manage to calculate the fraction $\dot{H}/H^2$ as a function of the variables of the dynamical system. As a result, we obtain an autonomous system and therefore, instead of making any assumptions based on different periods of the Universe, we will see in the following paragraphs that we can actually derive these periods from the critical points along with the value of the fraction.

In what follows, we keep the constants $k$ and $\lambda$ of the coupling function and the potential arbitrary. It turns out, that on every critical point, the fraction $\dot{H}/H^2$ is a constant. This means that from Eqs.~\eqref{eq:28}-\eqref{eq:32}, it is evident that when we focus on a critical point, the matter fields do not interact with the scalar field or the GB invariant. This indicates that Eqs.~\eqref{eq:28}-\eqref{eq:30} contain all the necessary information to comprehend the evolution of the cosmological model, whereas Eq.~\eqref{eq:31} and Eq.~\eqref{eq:32} evolve independently.

Let us proceed by calculating the critical points of the above systems in order to analyze their features and behavior. The physical properties and existence of the critical points are illustrated in Table~\ref{table:1}. Additionally, the hyperbolicity and stability evaluations are given in Table~\ref{table:2}. There are seven critical points and in the following, we will discuss the properties of each critical point separately along with their potential connection with various evolution eras of the Universe.

\begin{table}[H]\footnotesize
\hskip-0.2cm
\begin{tabular}{|M{0.9cm}||P{4.6cm}|M{1.5cm}|P{1.4cm}|P{1.4cm}|P{1.4cm}|P{1.4cm}|P{1.4cm}|P{1.4cm}|P{1.4cm}|} \hline
Point&$\{x_1,x_2,x_3,x_4,x_5\}$&\centering Existence&\centering$\Omega_{\phi}$&\centering$\Omega_{GB}$&\centering$\Omega_m$&\centering $\Omega_r$&\centering${\dot H}/H^2$&\centering$w_{\phi}$&$w_{\rm eff}$\\
\hline\hline
\centering A&$\{\,0,\,0,\,0,\,0,\,1\,\}$&Always&\centering$0$&\centering$0$&\centering$1$&\centering$0$&\centering$-3/2$&\centering$-$&$0$\\
\hline
\centering B&$\{\,0,\,0,\,0,\,1,\,0\,\}$&Always&\centering$0$&\centering$0$&\centering$0$&\centering$1$&\centering$-2$&\centering$-$&$1/3$\\
\hline
\centering C&$\{\,0,\,1,\,\lambda/8,\,0,\,0\,\}$&Always&\centering$1$&\centering$0$&\centering$0$&\centering$0$&\,\,\,$0$&\centering$-1$&$-1$\\
\hline
\centering D&$\{\,1,\,0,\,0,\,0,\,0\,\}$&Always&\centering$1$&\centering$0$&\centering$0$&\centering$0$&\centering$-3$&\,\,\,$1$&\,\,\,$1$\\
\hline
\centering E&$\{\,\dfrac{2 \sqrt{2/3}}{\lambda},\,\dfrac{2}{\sqrt{3} \lambda},0,\,\dfrac{\sqrt{\lambda^2-4}}{\lambda},\,0\,\}$&$\lambda^2>4$&$4/\lambda^2$&$0$&$0$&$\dfrac{\lambda^2-4}{\lambda^2}$&$-2$&$1/3$&$1/3$\\
\hline
\centering F&$\{\,\dfrac{\sqrt{3/2}}{\lambda},\,\dfrac{\sqrt{3/2}}{\lambda},\,0,\,0,\,\dfrac{\sqrt{\lambda^2-3}}{\lambda}\,\}$&$\lambda^2>3$&$3/\lambda^2$&$0$&$\dfrac{\lambda^2-3}{\lambda^2}$&$0$&$-3/2$&$0$&$0$\\
\hline
\centering G&$\{\,\,\dfrac{\lambda}{\sqrt{6}},\,\dfrac{\sqrt{6-\lambda^2}}{\sqrt{6}},0,0,0\,\}$&$\lambda^2<6$&$1$&$0$&$0$&$0$&$-\lambda^2/2$&$-1+\dfrac{\lambda^2}{3}$&$-1+\dfrac{\lambda^2}{3}$\\
\hline
\end{tabular}
\caption{Existence and physical properties of the critical points}\label{table:1}
\end{table}

\begin{table}[H]\footnotesize
\hskip-0.2cm
\begin{tabular}{|M{0.9cm}||P{8.2cm}|M{6cm}|P{2cm}|} \hline
Point&\centering Eigenvalues&\centering Hyperbolicity& Stability\\
\hline\hline
\centering A&$\{\,-3,\,3,\,-3/2,\,3/2,\,-1/2\}$&Hyperbolic&Saddle\\
\hline
\centering B&$\{\,-4,\,4,\,2,\,-1,\,1/2\}$&Hyperbolic&Saddle\\
\hline
\centering C&$\{\,-2,\,-3/2,\,0,\,\dfrac{1}{2} \,\left(\,-3\pm\sqrt{3}\sqrt{3-4 k\lambda-4\lambda^2}\,\,\right)\,\}$&Non-hyperbolic&Stable\\
\hline
\centering D&$\{\,6,\,3/2,\,1,\,-6+\textrm{k}\sqrt{6},\,\dfrac{1}{2}\,\left(6-\sqrt{6}\lambda\right)\,\}$&\begin{tabular}{m{2.3cm} m{3.3cm}}
 $\lambda$ or k$=\sqrt{6}$,& \,\,\,Non-hyperbolic \\[-0.5cm]
 $\lambda<\sqrt{6}$,\,k$>\sqrt{6}$,& \,\,\,\,Hyperbolic\\[-0.55cm]  $\lambda<\sqrt{6}$,\,k$<\sqrt{6}$,& \,\,\,\,Hyperbolic\\[-0.5cm]  $\lambda>\sqrt{6}$,\,k\,$\lessgtr{\sqrt{6}}$,&\,\,\,\,Hyperbolic\\
  \end{tabular}&\begin{tabular}{P{2cm}}
 Unstable \\[-0.5cm]
 Repeller\\[-0.55cm] 
 Saddle\\[-0.5cm] 
 Saddle
  \end{tabular}\\
\hline  
\centering E&$\{\,1/2,\,4,\,\dfrac{4\left(\textrm{k}\,\lambda-\lambda^2\right)}{\lambda^2},\,\pm \,\dfrac{-\lambda^2+\sqrt{-\lambda^2(-64+15\lambda^2)}}{2\lambda^2}\,\}$&\begin{tabular}{m{1.3cm} m{2.5cm}}
 $\lambda=\textrm{k}$,& \,\,\,Non-hyperbolic \\[-0.5cm]
 $\lambda\lessgtr{\textrm{k}}$,& \,\,\,\,Hyperbolic\\
  \end{tabular}&\begin{tabular}{P{2cm}}
 Unstable \\[-0.5cm]
 Saddle
  \end{tabular}\\
 \hline
 \centering F&$\{-1/2,\,3,\,\dfrac{3\left(\textrm{k}\,\lambda-\lambda^2\right)}{\lambda^2},\,\pm \,\dfrac{3(-\lambda^2+\sqrt{-\lambda^2(-24+7\lambda^2)})}{4\lambda^2}\}$&\begin{tabular}{m{1.3cm} m{2.5cm}}
 $\lambda=\textrm{k}$,& \,\,\,Non-hyperbolic \\[-0.5cm]
 $\lambda\lessgtr{\textrm{k}}$,& \,\,\,\,Hyperbolic\\
  \end{tabular}&\begin{tabular}{P{2cm}}
 Unstable \\[-0.5cm]
 Saddle
  \end{tabular}\\
  \hline
  \centering G&$\{\,\lambda^2,\,\textrm{k}\lambda-\lambda^2,\,\dfrac{1}{2}(-6+\lambda^2),\,\dfrac{1}{2} (-4+\lambda^2), \,\dfrac{1}{2} (-3+\lambda^2)\,\}$&\begin{tabular}{m{1.3cm} m{2.5cm}}
 $\lambda^2<2$&or\,\,\,\,\,\,$2<\lambda^2<6$, \\[-0.5cm]
 $\lambda=\textrm{k}$,& \,\,\,Non-hyperbolic \\[-0.5cm]
  $\lambda\lessgtr{\textrm{k}}$,& \,\,\,\,Hyperbolic\\
  \end{tabular}&\begin{tabular}{P{2cm}}
 \\[-0.5cm]
 Unstable\\[-0.55cm] 
 Saddle
  \end{tabular}\\
\hline  
\end{tabular}
\caption{Stability properties of the critical points}\label{table:2}
\end{table}

\begin{itemize}
    \item[$\circ$] \textbf{Point A.} The first critical point corresponds to matter dominated Universe. As it is evident from Table~\ref{table:1}, $\Omega_m=1$ and the critical point exists for all values of the free parameter $\lambda$. Additionally, the effective EoS matches the matter EoS, $w_{\rm eff}=w_m=0$ while the energy density of the scalar field is zero. As a result, there is no late-time acceleration for any physically accepted value of $w_{\rm eff}$. From Table~\ref{table:2}, it is obvious that we are dealing with a saddle hyperbolic critical point with a 3D local stable manifold where the trajectories approach point A moving tangently to the slower direction, that is the line spanned by the eigenvector corresponding to $-1/2$ eigenvalue. Furthermore, point A is characterized by a 2D local unstable manifold, where the description of local indicates that these manifolds with boundaries are defined only in the neighborhood of the critical point. 
    \item[$\circ$] \textbf{Point B.} The second critical point is related to a radiation dominated Universe. Here, $\Omega_r=1$ and the critical point exists for all values of the free parameter $\lambda$. Moreover, the effective EoS matches the radiation EoS, $w_{\rm eff}=w_r=1/3$ and the energy density of the scalar field is again zero. Similar to point A, point B does not exhibit late-time acceleration for any physically accepted value of $w_{\rm eff}$. As we can see from Table~\ref{table:2}, point B is a saddle hyperbolic critical point with 2D local stable manifold where the trajectories approach point B moving tangently to the slower direction, which is the line spanned by the eigenvector corresponding to $-1$ eigenvalue and a 3D local unstable manifold. Again, the term local implies manifolds defined in the vicinity of point B.
    \item[$\circ$] \textbf{Point C.} It is obvious from Table~\ref{table:1}  that the third critical point is completely dominated by the potential energy of the scalar field which behaves as a cosmological constant. With $w_{\rm eff}=w_{\phi}=-1$, the Universe is driven towards acceleration while matter and radiation densities are suppressed by that of the scalar field which dominates point C. Since the coupling to GB invariant does not vanish, point C might be related to early or late-time accelerated expansion phase of the Universe. Now, from Table~\ref{table:2} we notice that point C is a non-hyperbolic critical point characterized by one center manifold and a 4D stable manifold so linear stability theory cannot determine its behavior. By choosing to ignore matter and radiation and considering $3-4k\lambda-4{\lambda}^2>0$, the remaining equations are investigated using center manifold theory and the dimensionality of the system is reduced to one. In the reduced system, point C is characterized as stable critical point. Since the Hubble rate is constant, $\dot{H}=0,$ in the vicinity of the critical point, the expansion remains accelerating. Therefore, the third critical point behaves as the final attractor of the dynamical system.
    \item[$\circ$] \textbf{Point D.} The fourth critical point describes a Universe totally dominated by the kinetic energy of the scalar field. From Table~\ref{table:1} we can observe that $\Omega_{\phi}=1$ and the critical point exists for all values of the free parameters k and $\lambda$. As $w_{\rm eff}=w_{\phi}=1$, it is obvious that point D does not allow the possibility of acceleration. Moreover, since $\rho_{\phi}\sim
    {\alpha}^{-6}$, the energy density of the scalar field decreases much faster than the background density. From Table~\ref{table:2} it is evident that point D is unstable or saddle critical point depending on the values of k and $\lambda$ being greater or smaller than $\sqrt{6}$. Gathering all this information, we conclude that critical points with steep potential such as point D cannot be associated with any of the evolution periods of the CDM model or the late-time accelerated expansion. Nonetheless, these solutions are most likely associated with the early stages of evolution of the Universe.
    \item[$\circ$] \textbf{Point E.} The characteristic feature of the fifth critical point is that it describes a scaling solution. From Table~\ref{table:1} we notice that the fraction $\Omega_{\phi}/\Omega_r=4/\lambda^2-4$ is a non-zero constant implying a certain period where both the radiation and the scalar field could effect the evolution of the Universe. When this critical point exists, i.e. $\lambda^2>4$, the effective EoS matches the radiation EoS, $w_{\rm eff}=w_r=1/3$ and $w_{\phi}=w_r$, therefore point E cannot exhibit acceleration and the Universe evolves as if it was radiation dominated. During that period of evolution the presence of the scalar field could be hidden on cosmological scales. From Table~\ref{table:2} we can observe that the overall stability behavior depends on the relation between the parameters k and $\lambda$ . Nonetheless, point E is not a stable critical point in any case meaning that we do not need an extra mechanism to exit the scaling regime as is the case, for example, in the usual quintessence models.
    \item[$\circ$] \textbf{Point F.} The sixth critical point describes again a scaling solution. To be more accurate, point F features a matter scaling solution where the effective EoS matches the matter EoS, $w_{\rm eff}=w_m=o$ and $w_{\phi}=w_m$ while the fraction $\Omega_{\phi}/\Omega_m=3/\lambda^2-3$ is a non-zero constant. When this point exists,  i.e. $\lambda^2>3$, the Universe evolves as if it was matter dominated meaning that point F cannot exhibit acceleration either. From Table~\ref{table:2} it is obvious that its stability behaviour depends on the relation between parameters k and $\lambda$ . Similar to point E, point F is not a stable point either.
    \item[$\circ$] \textbf{Point G.} From Table~\ref{table:1} it is evident that the seventh critical point exists for $\lambda^2<6$. As $w_{\rm eff}=w_{\phi}=-1+\lambda^2/3$, we recognize that point G is a scalar field dominated critical point. Therefore, to allow the possibility of acceleration, the condition $w_{\rm eff}<-1/3$ must hold which means that $\lambda^2<2$ must be satisfied. At this point it is interesting to note that in the limit $\lambda \to 0$ the EoS of cosmological constant is recovered which  indicates the stable accelerated point C of our study. Away from the limit $\lambda \to 0$ and as long as $\lambda^2<2$, we can see from  Table~\ref{table:2} that point G is not at all a stable accelerated point. Depending on the relation between the parameters k and $\lambda$, we are dealing with a non-hyperbolic or hyperbolic critical point, which nonetheless is not stable and cannot account for the final attractor of the system. Searching for  a valid interpretation of  point G, one reasonable option is to link it with an early accelerated period of the evolution of the Universe. To be precise, point G could possibly account for the inflationary era of the system since for $\lambda^2<2$ we have that ${\phi}^2 \ll V(\phi) $, so the potential is shallow enough to allow such a period to be emerged in the early Universe.
\end{itemize}

According to the previous analysis of the critical points and provided that $3-4k\lambda-4{\lambda}^2>0$, it is obvious that only point C could eventually give rise to a stable accelerated attractor of the system under study. Bearing that in mind, the radiation-dominated era could be realized by either point B or the scaling solution described by point E. However, as previously stated, point E exists only for $\lambda^2>4$ which contradicts the assumption of $3-4k\lambda-4{\lambda}^2>0$ for the point C. As a result, radiation dominated era is realized by point B. Regarding the matter-dominated era, there are two options as well, namely point A or the scaling solution described by point F. In this case, the condition $\lambda^2>3$ must hold in order for point F to exist which again is opposed to the assumption $3-4k\lambda-4{\lambda}^2>0$ of point C. Therefore, point A is the only way to realize matter-dominated era.

\section{Visualization of Phase Portraits} \label{sec:phase_port}

The description of each critical point's behavior through proper diagrams is an effective strategy for fully comprehending their distinguishing features. In this section we introduce the phase portraits for each critical point, highlight the crucial steps for their derivation and discuss whether they are compatible with the analysis of Sec.~\ref{sec:dyn_sys_ana}.

\begin{figure}[h]
\includegraphics[width=6.5cm, height=6.5cm]{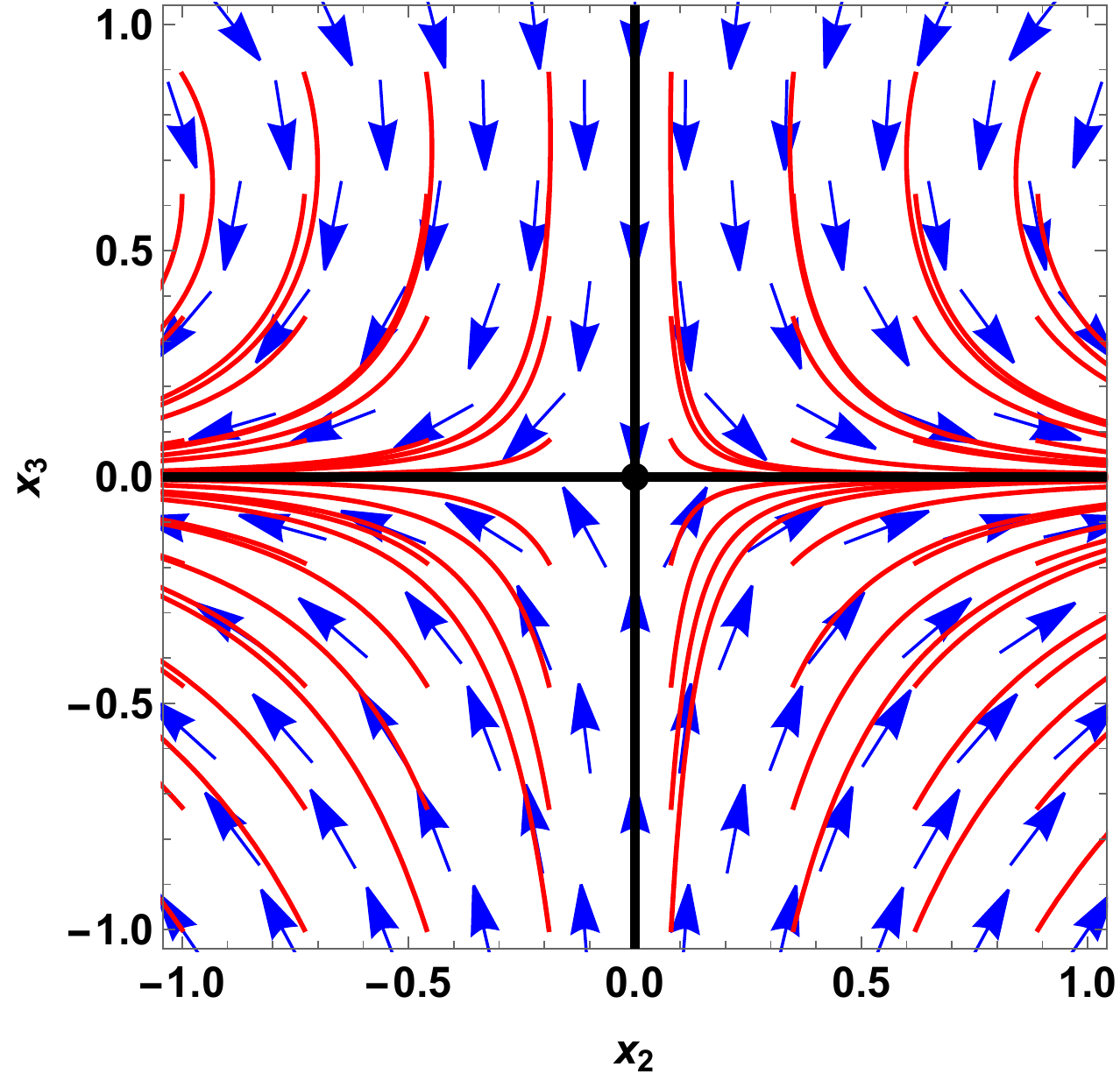}
\caption{2D phase portrait of point A.}
\label{fig:imA}
\end{figure}

Based on the analysis of the previous section, point A is a matter dominated solution of the model which exists for all values of the free parameters. Obviously, point A cannot represent accelerated behavior for any physically accepted value of $w_{\rm eff}$ since $w_{\rm eff}=w_m=0$.

As $x_1=4\sqrt{6}\,x_3/3+\sqrt{3}\,x_2^2/3\,\sqrt{2}$  on that critical point, we ended up with two equations for the variables $x_2,x_3$, namely  $x_2'=3\,x_2/2-x_2^3/4-2x_2\,x_3$ and $x_3'=-3\,x_3+x_2^2\,x_3/4+2\,x_3^2$, for  $k=1/4$ and $\lambda=1/2$. Figure~\ref{fig:imA} displays the two-dimensional phase portrait of point A regarding $x_2,x_3$ and for  $k=1/4$ and $\lambda=1/2$. Blue arrows indicate the vector field while red lines are numerical solutions representing some of the trajectories in the vicinity of point A. Here, the critical point is defined by the black dot in the center of the diagram. According to Tables~\ref{table:1}, \ref{table:2}, point A is hyperbolic with eigenvalues $\,-3,\,3,\,-3/2,\,3/2,\,-1/2$. Therefore it is a saddle critical point regardless of the choice of the free parameters $k$ and $\lambda$. The phase portrait of Figure~\ref{fig:imA} visualises the saddle behavior of point A and verifies the conclusions of Tables~\ref{table:1}, \ref{table:2}.

\begin{figure}[h]
\includegraphics[width=6.5cm, height=6.5cm]{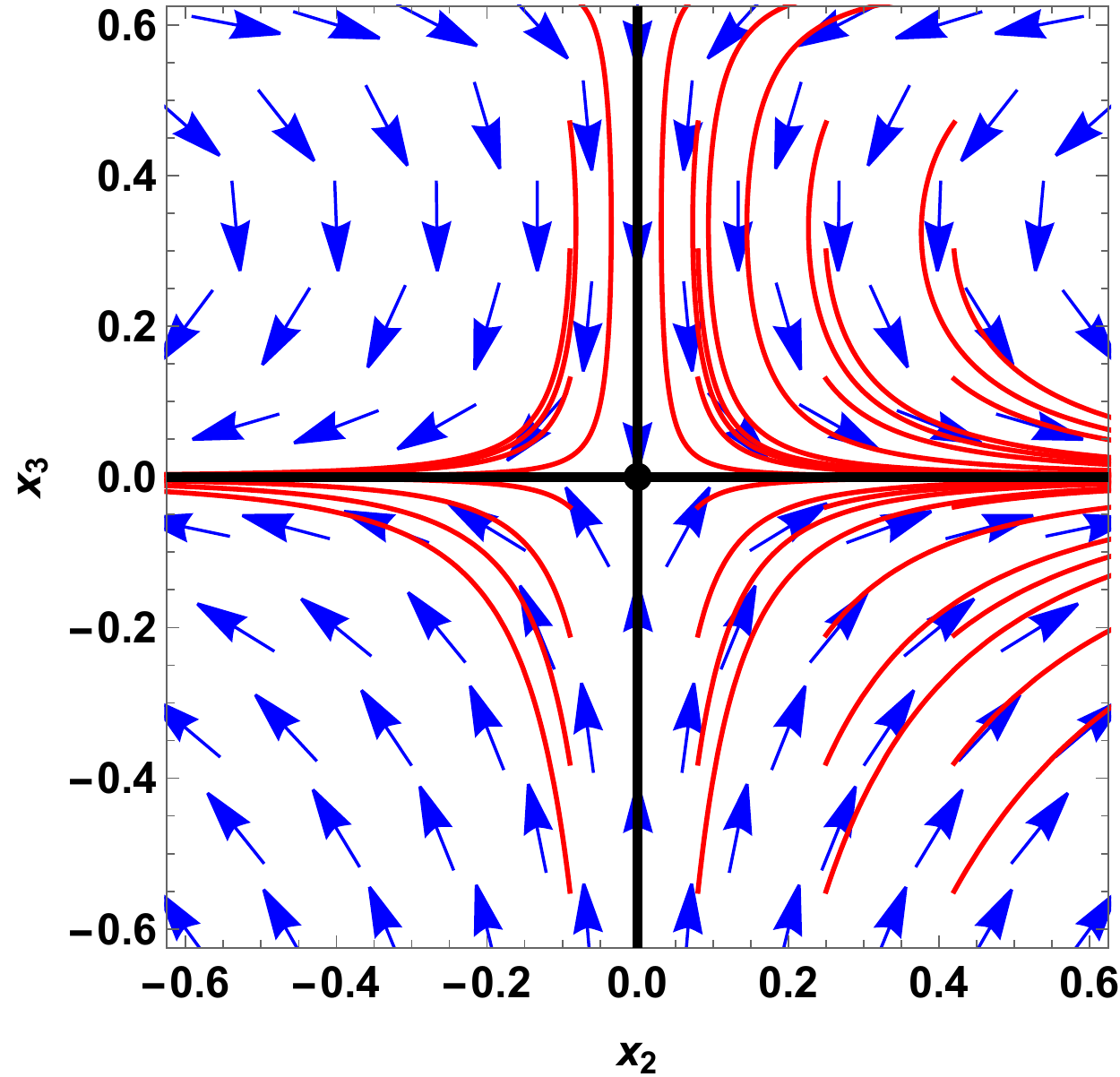}
\caption{2D phase portrait of point B.}
\label{fig:imB}
\end{figure}

The second critical point, which exists for all values of the free parameters, indicates a Universe dominated by radiation. Just like point A, point B cannot account for an accelerated solution for any physically accepted value of $w_{\rm eff}$, since in this case $w_{\rm eff}=w_r=1/3$.

As $x_1=4\sqrt{6}\,x_3\,+\sqrt{3}\,x_2^2/2\,\sqrt{2}$ on the critical point, the equations for the variables $x_2,x_3$ are $x_2'=2\,x_2-3\,x_2^3/8-6\,x_2\,x_3$ and $x_3'=-4\,x_3+3\,x_2^2\,x_3/8+6\,x_3^2$ for  $k=1/4$ and $\lambda=1/2$. Figure~\ref{fig:imB} displays the two-dimensional phase portrait of point B regarding $x_2,x_3$ and for $k=1/4$ and $\lambda=1/2$. The vector field is defined by blue arrows and red lines are numerical solutions representing some of the trajectories in the vicinity of point B. Again, the critical point is defined by the black dot in the center of the diagram. As stated in the Tables~\ref{table:1}, \ref{table:2}, point B is hyperbolic with eigenvalues $\,-4,\,4,\,2,\,-1,\,1/2$. Therefore it is a saddle critical point regardless of the choice of the free parameters $k$ and $\lambda$. Figure~\ref{fig:imB} visualises the saddle behavior of point B and verifies the results of Table~\ref{table:2}.

\begin{figure}[h]
\includegraphics[width=6.5cm, height=6.0cm]{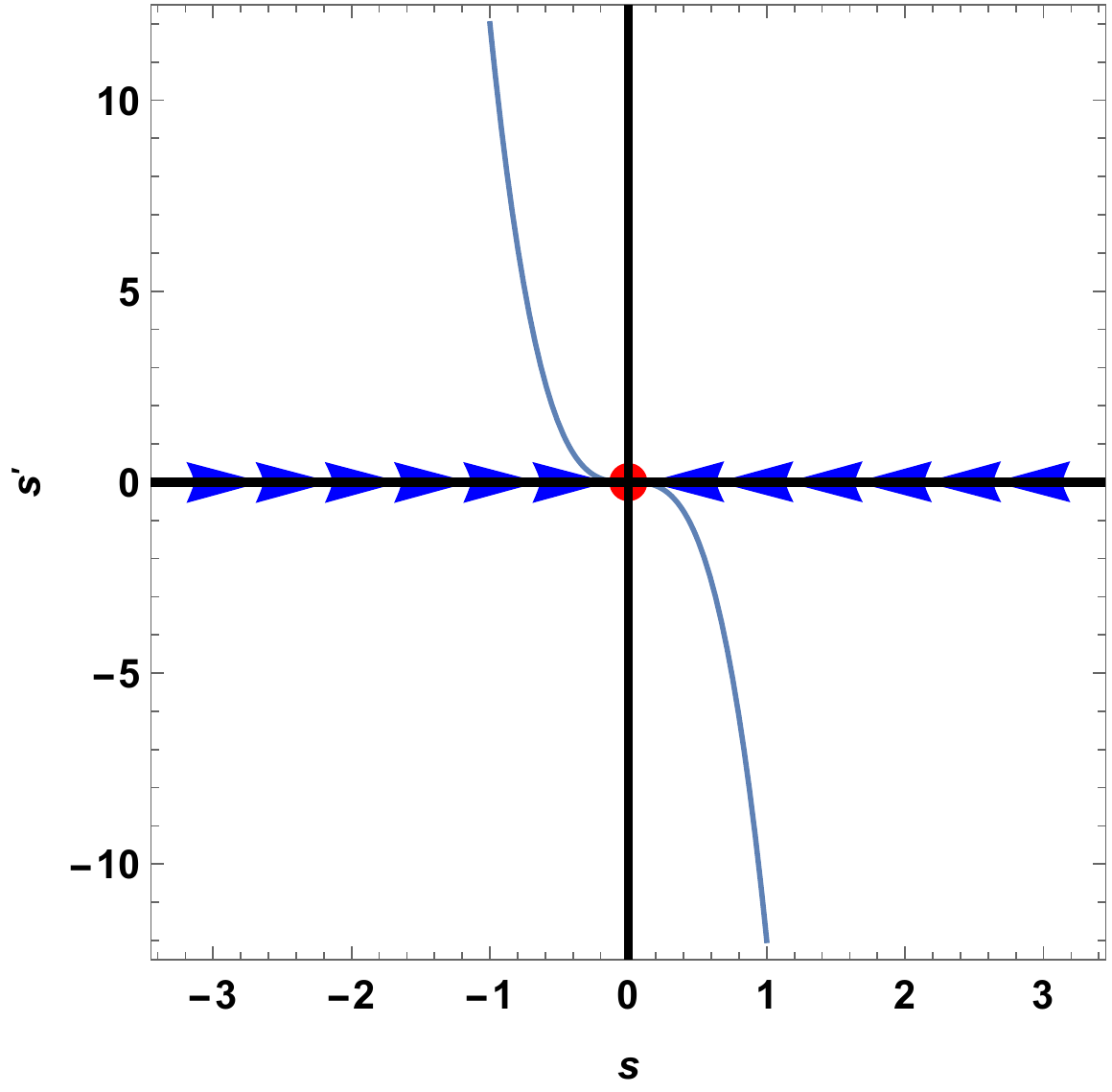}
\caption{1D phase portrait of point C.}
\label{fig:imC}
\end{figure}

Following Table~\ref{table:1} the third critical point is dominated by the potential energy $V(\phi)$ of the scalar field. Since $w_{\rm eff}=w_{\phi}=-1$, point C could account for the accelerated solution of the model. From its eigenvalues listed in Table~\ref{table:2} it is obvious that point C is non-hyperbolic, so linear stability theory cannot determine its behavior. The phase space of the new variables X,Y,Z  is obtained by a shift to the origin according to $X=x_1,Y=x_2-1,Z=x_3-\lambda/8$ and for $\lambda=k=1/2$. By  applying center manifold theory and ignoring matter and radiation, the dimensions of the system are reduced to one. The behavior of the new variable $s(t)$ is determined by the differential equation $s'(t)=s(t)^3 \left(-12k/\lambda (k+\lambda)^2\right)$. Figure~\ref{fig:imC} displays its one-dimensional phase portrait. The grey line represents the trajectory for $\lambda=k=1/2$ while blue arrows are for the vector field.  Point C, defined by the red dot in the center of the diagram, is a stable critical point acting as the final attractor of the model.

\begin{figure}[h]
\includegraphics[scale=0.7]{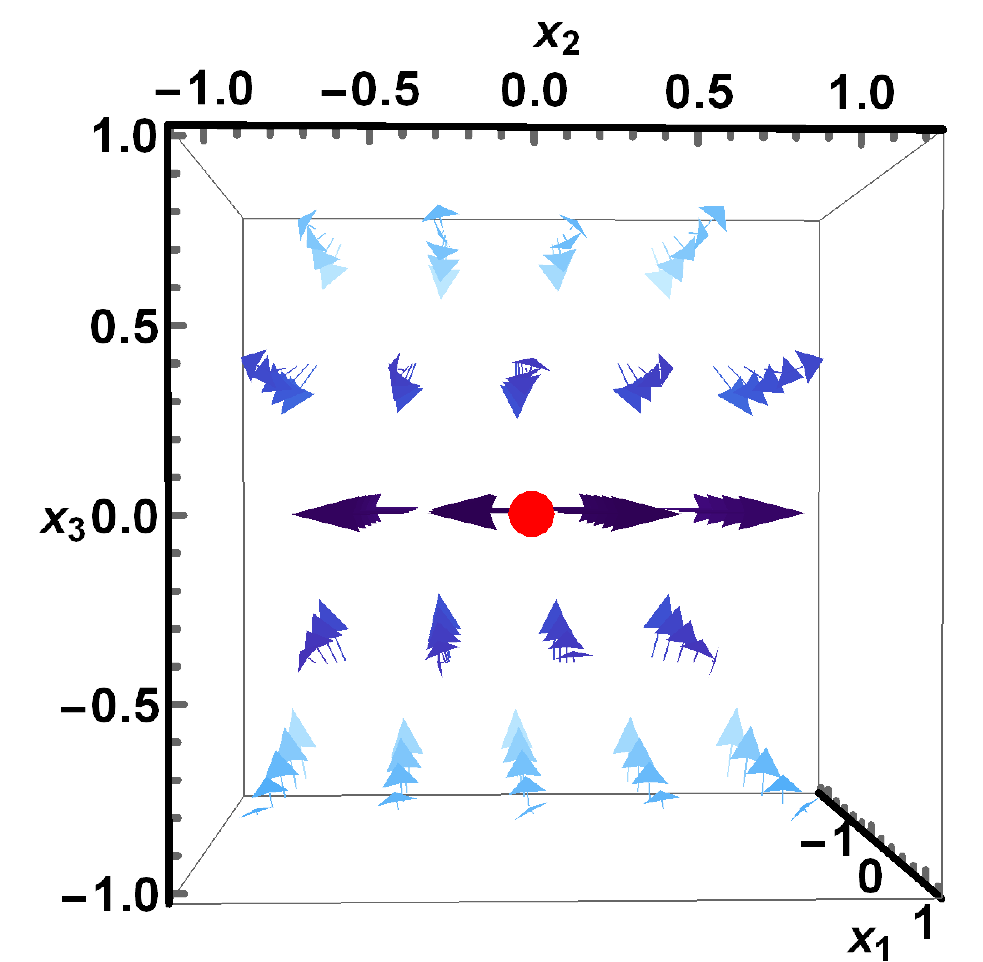}
\caption{3D vector plot of point D.}
\label{fig:imD}
\end{figure}

The fourth critical point defines a Universe dominated by the kinetic energy of the scalar field. Since $w_{\rm eff}=w_{\phi}=1$, it is obvious that point D does not allow acceleration. Following Table~\ref{table:2}, the behavior of point D- unstable or saddle- depends on the values of $k$ and $\lambda$ being greater or smaller than $\sqrt{6}$ although point D exists regardless of the values of the parameters. Figure~\ref{fig:imD} demonstrates the three-dimensional vector plot of $x_1,x_2,x_3$ for $k=3$ and $\lambda=1$. Blue arrows indicate the vector field in the vicinity of the critical point, and the red dot defines its coordinates at $x_1=1,x_2=0,x_3=0$. According to Table~\ref{table:2}, and for that particular choice of the parameters $k$, $\lambda$, point D is hyperbolic and behaves as a repeller of the model.

\begin{figure}[h]
\includegraphics[width=6.5cm, height=6.5cm]{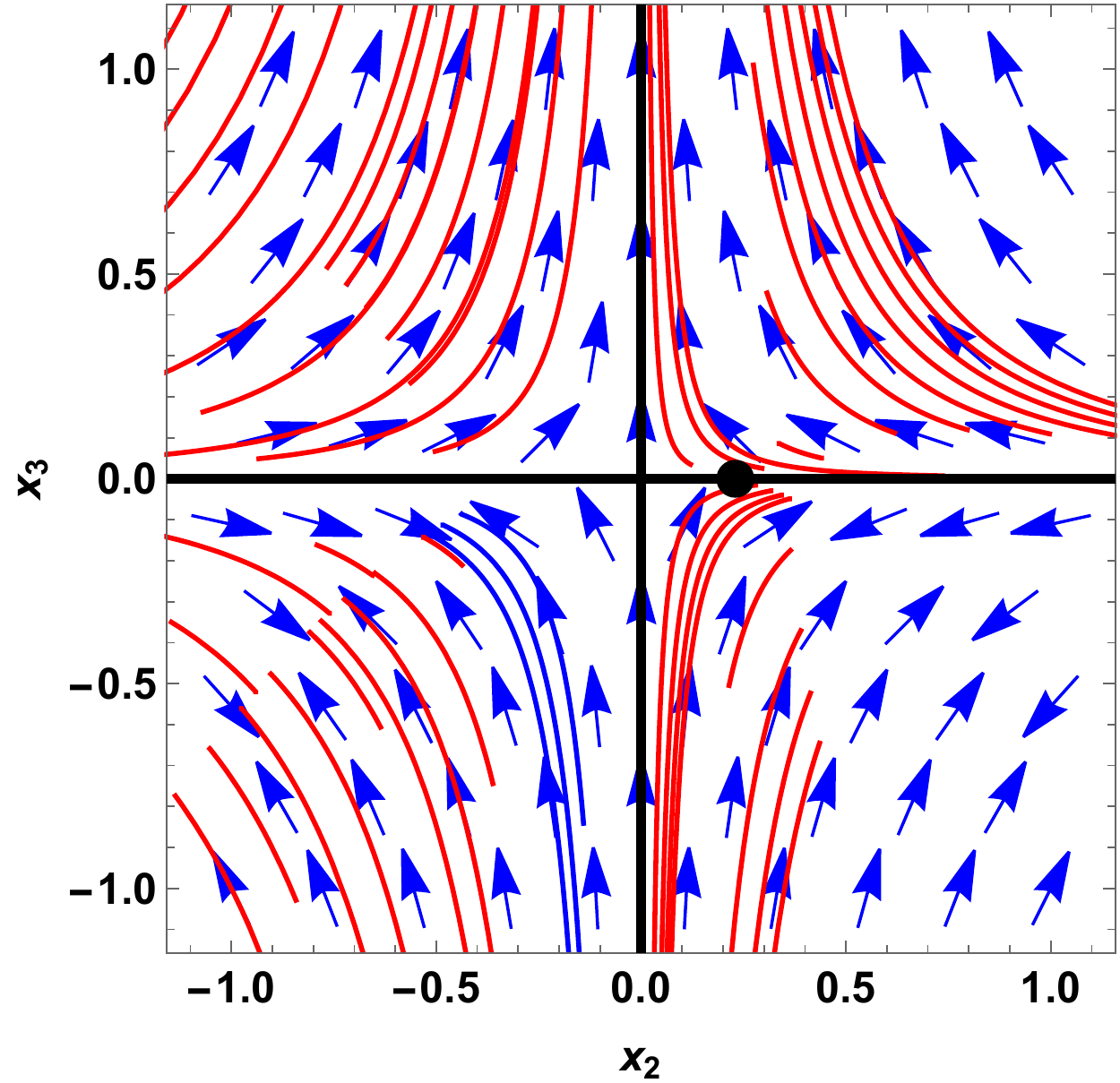}
\caption{2D phase portrait of point E.}
\label{fig:imE}
\end{figure}

Point E exists for $\lambda^2>4$ and when it does, it represents a radiation scaling solution of the system with $\Omega_{\phi}/\Omega_r=4/\lambda^2-4$ a non-zero constant. Since $w_{\rm eff}=w_r=1/3$ and $w_{\phi}=w_r$, point E cannot stand for an accelerated solution.

As $x_1=4\sqrt{6}\,x_3+5\,\sqrt{3/2}\,x_2^2$ on the critical point, the remaining variables $x_2,x_3$ behave as $x_2'=2\,x_2-75\,x_2^3/2-60\,x_2\,x_3$ and $x_3'=-4\,x_3+75\,x_2^2\,x_3+120\,x_3^2$ for $k=5$ and $\lambda=5$. Figure~\ref{fig:imE} displays the two-dimensional phase portrait of point E regarding $x_2,x_3$. Blue arrows denote again the vector field and red lines are numerical solutions representing some of the trajectories in the neighborhood of point E. The critical point is defined by the black dot with coordinates $x_2\simeq0.23,x_3=0$. According to the Tables~\ref{table:1}, \ref{table:2}, and for $k=5$ and $\lambda=5$, point E exists, is a scaling solution and it behaves as an unstable non-hyperbolic critical point.

\begin{figure}[h]
\includegraphics[width=6.5cm, height=6.5cm]{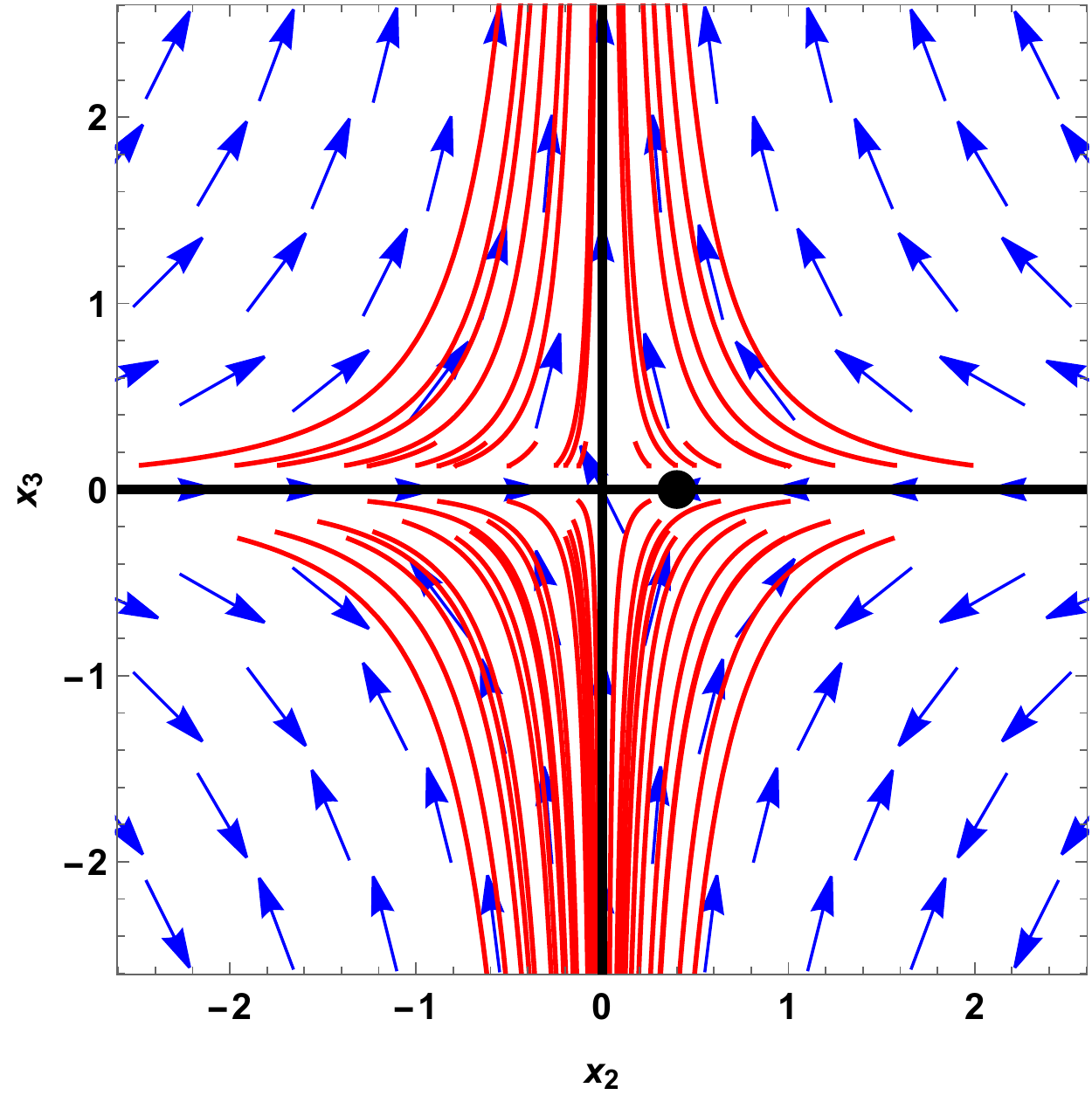}
\caption{2D phase portrait of point F}
\label{fig:imF}
\end{figure}

As stated in the previous section, point F exists for $\lambda^2>3$ and in that case it describes a matter scaling solution with $\Omega_{\phi}/\Omega_m=3/\lambda^2-3$ a non-zero constant. For this scaling solution $w_{\rm eff}=w_m=0$ and $w_{\phi}=w_m$, therefore point F cannot exhibit acceleration either.

As $x_1=4\sqrt{2}\,x_3/\sqrt{3}+\,\sqrt{6}\,x_2^2$ on the critical point, the differential equations describing the behavior of the variables $x_2,x_3$ are $x_2'=3\,x_2/2-9\,x_2^3-12\,x_2\,x_3$ and $x_3'=-3\,x_3+18\,x_2^2\,x_3+24\,x_3^2$ for $k=4$ and $\lambda=3$. Figure~\ref{fig:imF} displays the two-dimensional phase portrait of the second scaling solution, for $k=4$ and $\lambda=3$. Blue arrows denote the vector field and red lines are numerical solutions representing some of the trajectories in the vicinity of point F. The critical point is defined by the black dot at $x_2\simeq0.4,x_3=0$. According to Tables~\ref{table:1},\ref{table:2}, and for $k=4$ and $\lambda=3$, point F exists and it behaves as a saddle hyperbolic critical point.

\begin{figure}[h]
\includegraphics[width=6.5cm, height=6.5cm]{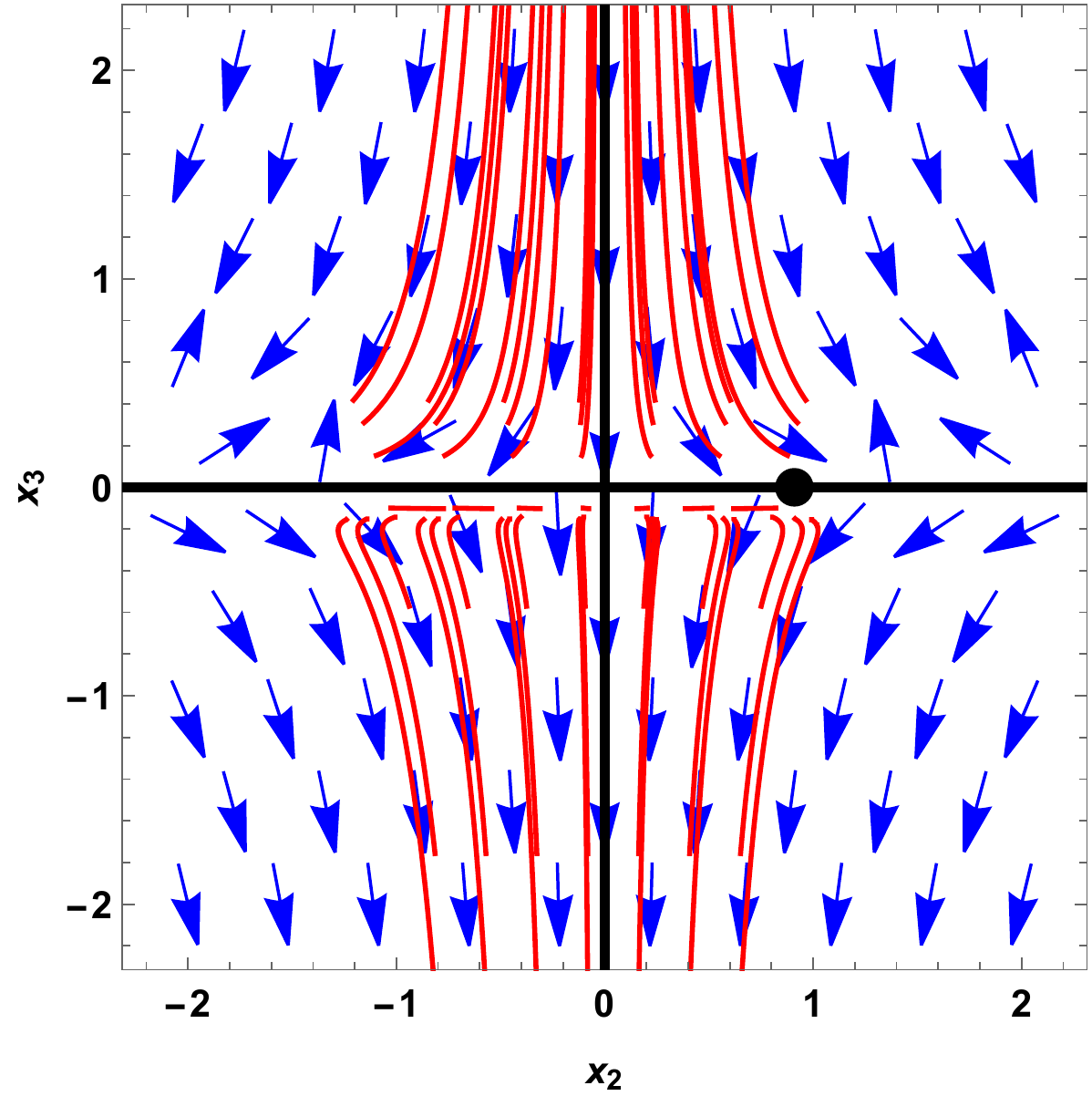}
\caption{2D phase portrait of point G.}
\label{fig:imG}
\end{figure}

The last critical point exists for $\lambda^2<6$ and when it does, it describes a scalar field dominated Universe. Since $w_{\rm eff}=w_{\phi}=-1+\lambda^2/3$, the condition $\lambda^2<2$ must be satisfied in order for point G to be an accelerated solution. 

As $x_1=-4\sqrt{6}\,x_3/5+\,\sqrt{6}\,x_2^2/5$ on the critical point, the differential equations describing the behavior of the variables $x_2,x_3$ are $x_2'=x_2/2-3\,x_2^3/5+12\,x_2\,x_3/5$ and $x_3'=-x_3+12\,x_2^2\,x_3/5-48\,x_3^2/5$ for $k=2$ and $\lambda=1$. Figure~\ref{fig:imG} displays the two-dimensional phase portrait of the solution described by point G regarding $x_2,x_3$,
for  $k=2$ and $\lambda=1$. Because of the fact that $\lambda^2=1<2$, point G could represent an  accelerating phase of the model. Just as before, blue arrows denote the vector field and red lines are numerical solutions representing some of the trajectories in the vicinity of point G. The critical point is defined by the black dot with coordinates $x_2\simeq0.91,x_3=0$. As stated in Table~\ref{table:2} and for that particular choice of the parameters, point G is hyperbolic, accelerated but nonetheless saddle critical point.

\section{Conclusions} \label{sec:con}

A number of methods for performing a dynamical system analysis of the EGB theory are already present in the literature. Within this framework, our approach to conducting a similar analysis has a different orientation. The most important aspect of this work, is the impact on the dynamical system when the EGB theory is required to be compatible with the gravitational wave-speed constraint. To be more specific, as discussed in Sec.~\ref{sec:dyn_sys_ana}, the compatibility requirement described by Eq.~\eqref{eq:26} results in Eq.~\eqref{eq:27}. In this way we were able to express the fraction $\dot{H}/H^2$ as a function of the variables of the dynamical system. Following this new element, it was found that the autonomous feature of the dynamical system could be established from the start without making any assumptions concerning the value of that particular fraction over various cosmological eras. As a result, we derived these periods from the characteristic features and behavior of the critical points along with the value of the fraction.

The critical points of the autonomous dynamical system are seven in our case, which led to a rather compelling and  rich phenomenology of the EGB theory. According to the analysis, it is important to note that two of them could be related to early evolutionary eras of the Universe. For example, it was found that point G behaves as an unstable and accelerated solution as long as $\lambda^2<2$. This characteristic feature of the seventh critical point could possibly be related to the inflationary era of the model. In the same trend, it is found that in point D $w_\phi$ is driven to its maximum value, $w_\phi \rightarrow 1$, ruling out the possibility of acceleration. Since this corresponds to a stage in which the kinetic energy of the scalar field is dominant, the field rolls down the potential quickly. Thereby, point D is unlikely to be related to any of the different periods of the CDM model. All of these research findings indicate that points D and G merit further investigation regarding their potential correlation with the early stages of the Universe.

On the other hand, point C is completely dominated by the potential energy of the scalar field which behaves as a cosmological constant according to $w_\phi =- 1$. This means that the Universe is driven towards acceleration and because point C is found to be a stable critical point, it may account for the final attractor of the model.\par
The critical points analysis provided us with two scaling solutions namely points E and F where the energy density of the scalar field tracks that of the background perfect fluid.  Despite their initially prominent features due to their role in model building for dark energy, these two scaling solutions could not be combined with point C in order to realize a valid evolutionary sequence of the Universe. As a result and in order to observe matter and radiation dominated eras we employ points A and B of the dynamical system analysis.

\appendix*
\section{3-dimensional plots of the critical points}

This section outlines the three-dimensional vector plots for the critical points of the dynamical system under study. Additionally, it contains comments and a brief description regarding their characteristic features. Each three-dimensional plot could provide a more comprehensive perspective of the behavior of the critical points in accordance with  their analysis in Section~\ref{sec:dyn_sys_ana}  and the two-dimensional plots of Section~\ref{sec:phase_port}.

\begin{figure}[h]
     \begin{subfigure}[h]{0.3\textwidth}
         \includegraphics[width=5cm, height=5cm]{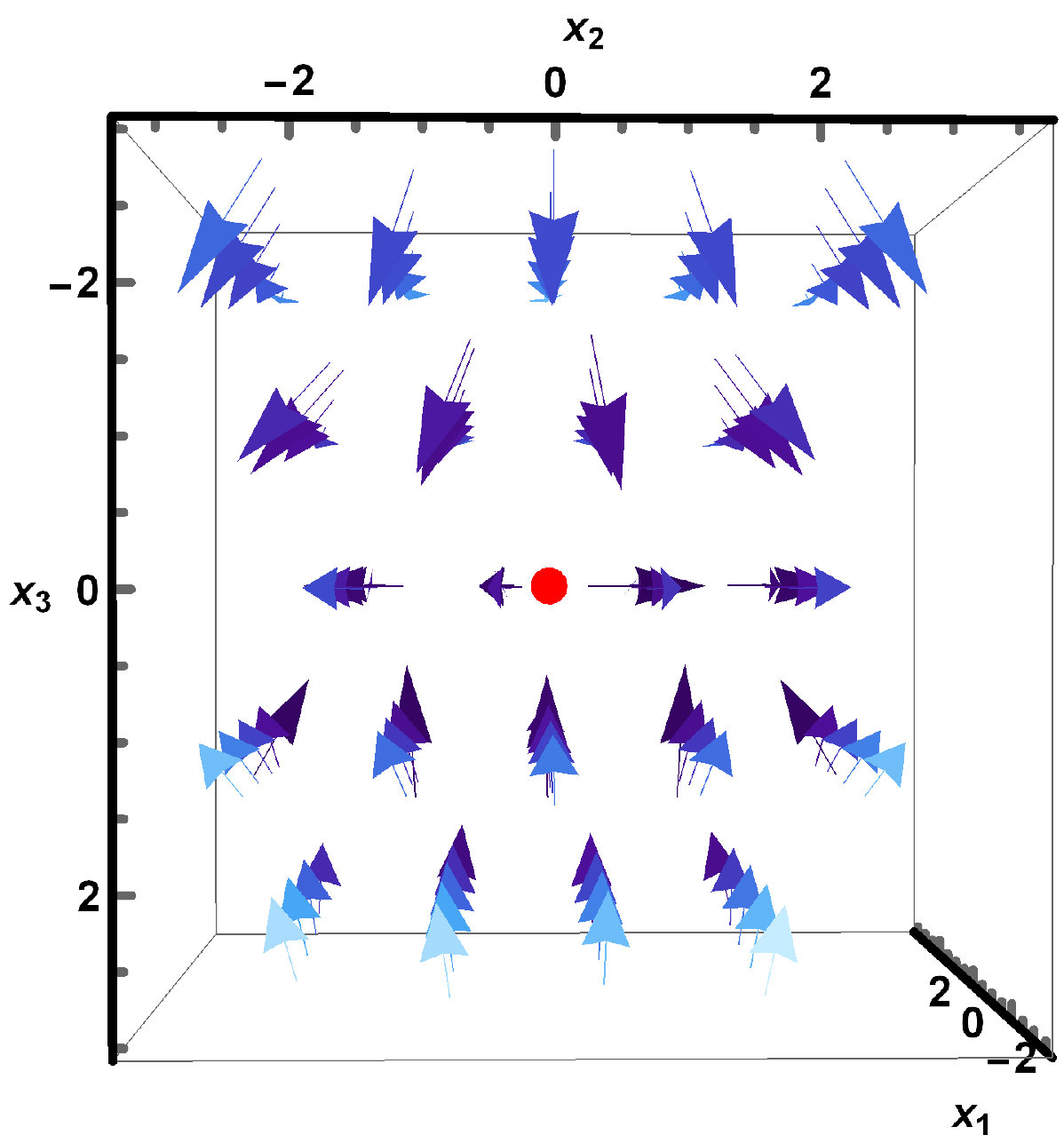}
%         \label{fig:A1}
     \end{subfigure}
     \begin{subfigure}[h]{0.3\textwidth}
         %\centering
         \includegraphics[width=5cm, height=5cm]{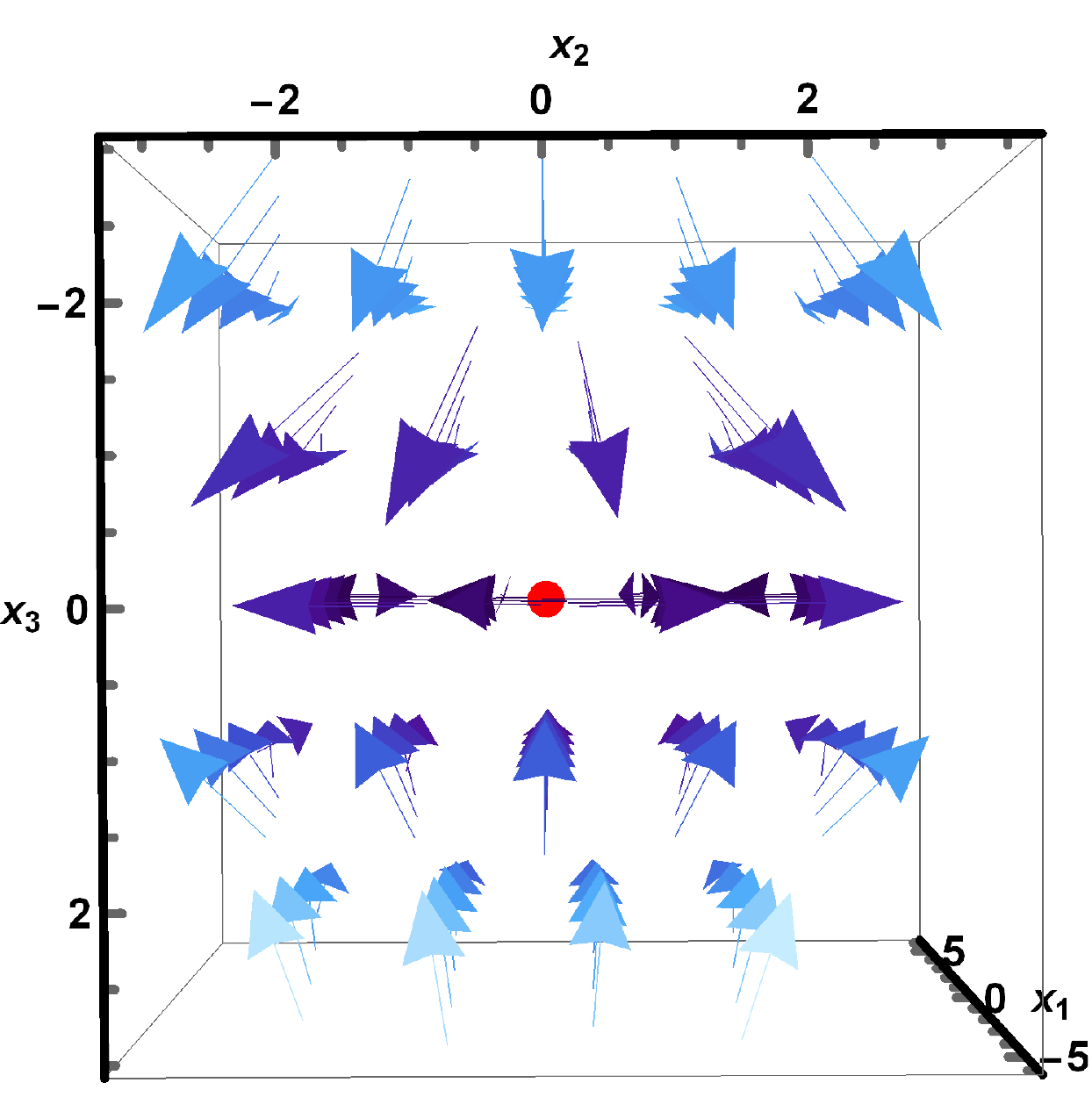}
%         \label{fig:B1}
     \end{subfigure}
     \begin{subfigure}[h]{0.3\textwidth}
         %\centering
         \includegraphics[width=5cm, height=5cm]{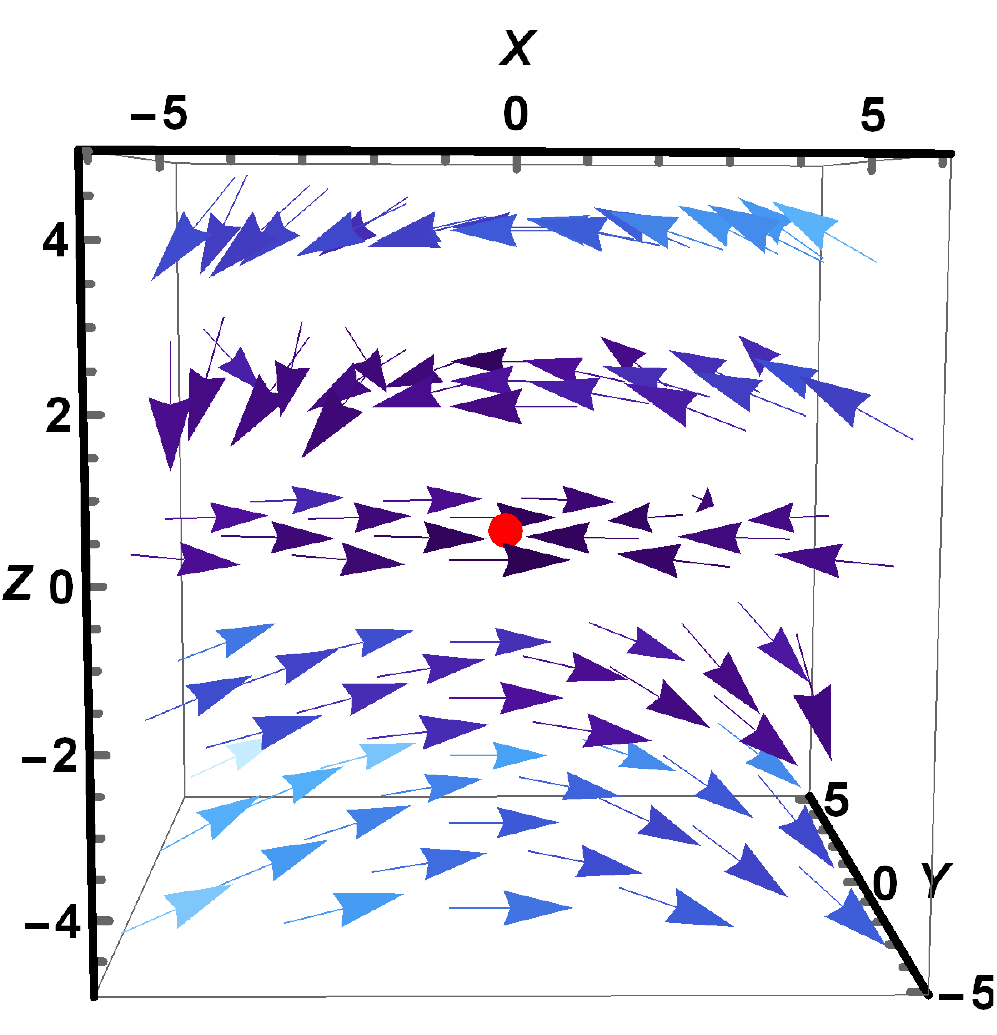}
%         \label{fig:B1}
     \end{subfigure}
        \caption{3D vector plots of points A, B and C.}
        \label{fig:A1B1C1}
\end{figure}

The left figure in Fig.~\ref{fig:A1B1C1} demonstrates the three-dimensional phase space of $x_1,x_2,x_3$ for $k=1/4$ and $\lambda=1/2$. Blue arrows denote the vector field while point A is defined by a red dot in the center of the plot. The vector plot clearly indicates that the defining feature of point A is that it is a saddle critical point of the model, just as was expected. The figure in the middle displays the three-dimensional phase space of $x_1,x_2,x_3$ for $k=1/4$ and $\lambda=1/2$. Again blue arrows denote the vector field and point B is defined by a red dot in the center of the plot. As predicted, the radiation dominated solution described by point B behaves as a saddle critical point.

The right figure in Fig~\ref{fig:A1B1C1} displays the phase space of X,Y,Z obtained by a shift to the origin of $x_1,x_2,x_3$ according to $X=x_1,Y=x_2-1,Z=x_3-\lambda/8$ and for $\lambda=k=1/2$. Again blue arrows denote the vector field while the critical point is defined by a red dot in the center of the plot. It is obvious that point C, which represents an accelerating solution of the model, is a stable critical point in accordance with Figure~\ref{fig:imC} and Table~\ref{table:2}.

\begin{figure}[h]
     \begin{subfigure}[h]{0.3\textwidth}
         \includegraphics[width=5cm, height=5cm]{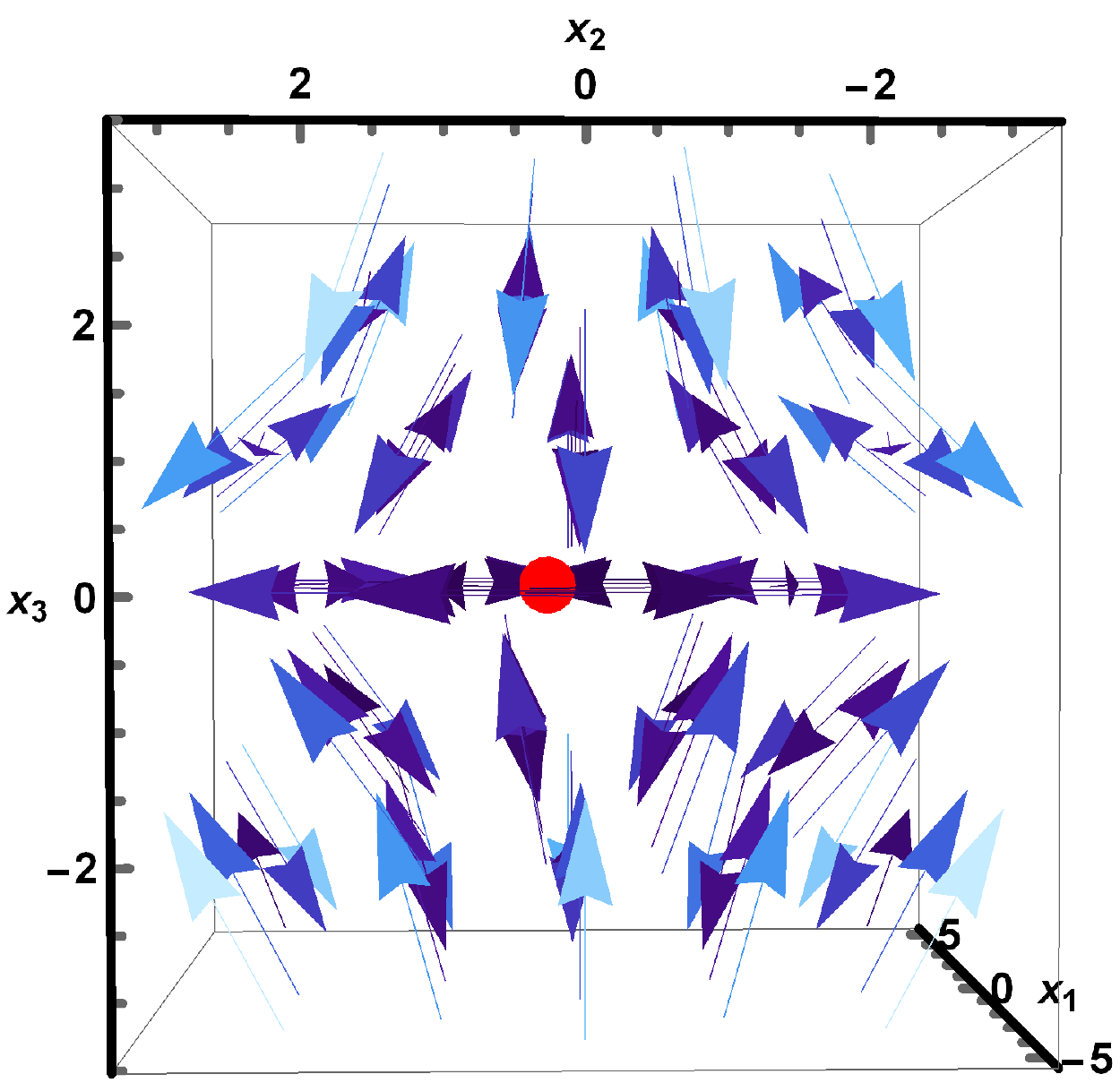}
     \end{subfigure}
     \begin{subfigure}[h]{0.3\textwidth}
         \includegraphics[width=5cm, height=5cm]{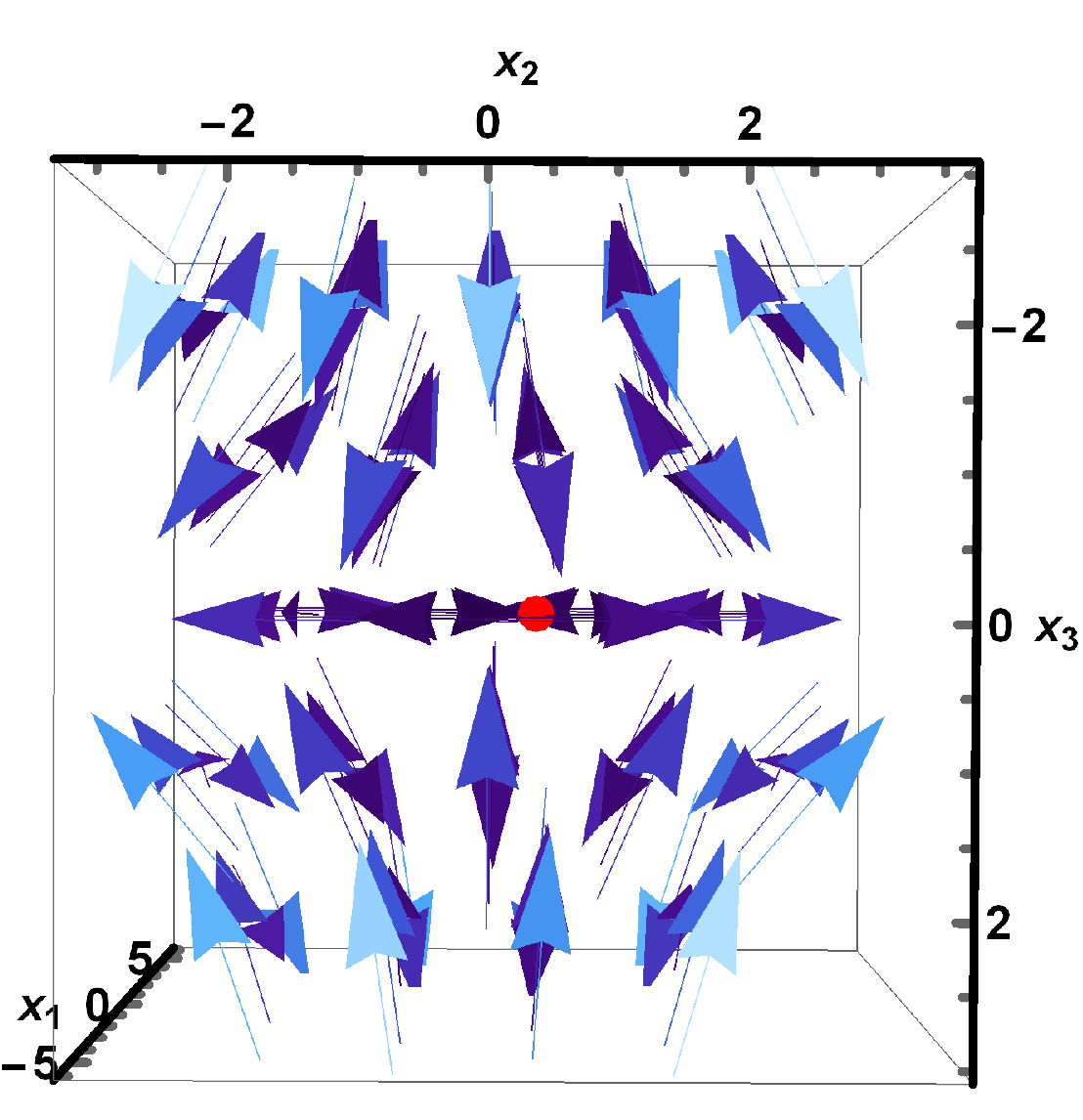}
     \end{subfigure}
     \begin{subfigure}[h]{0.3\textwidth}
        \includegraphics[width=5cm, height=5cm]{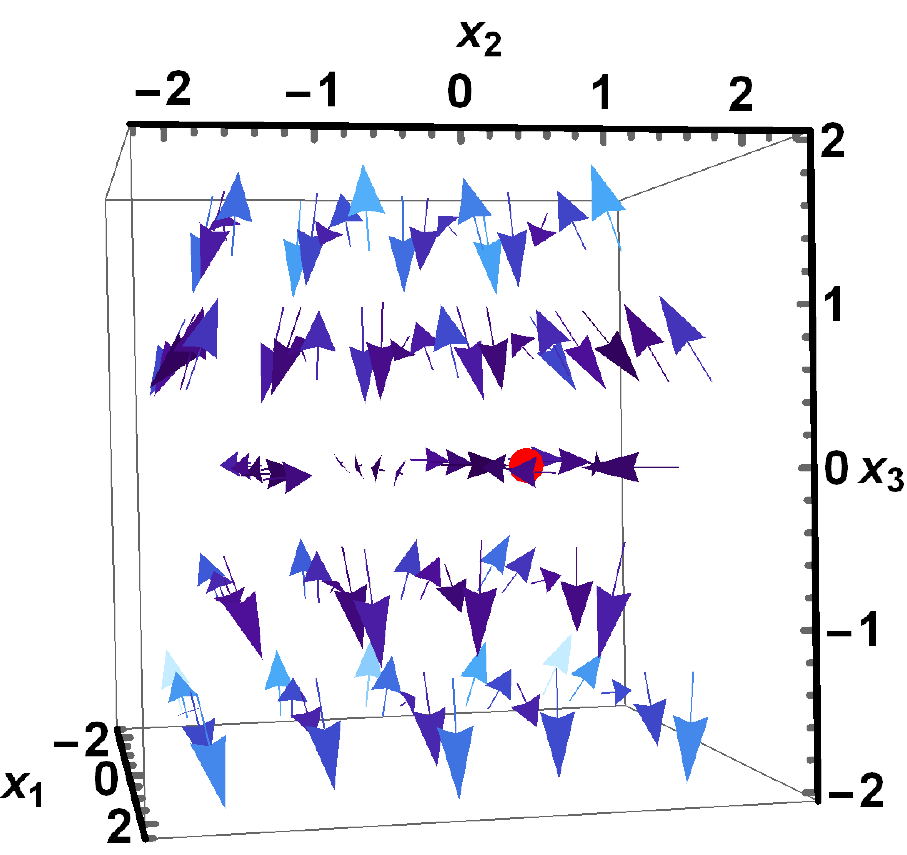}
     \end{subfigure}
        \caption{3D vector plots of points E, F and G.}
        \label{fig:E1F1G1}
\end{figure}

As explained earlier, point E is one of the two scaling solutions of the model. Figure~\ref{fig:E1F1G1} displays the three-dimensional phase space of $x_1,x_2,x_3$ for $k=5$ and $\lambda=5$. Blue arrows denote the vector field while the critical point is defined by a red dot with coordinates $x_1\simeq0.33,x_2\simeq0.23,x_3=0$. According to the Tables~\ref{table:1} and \ref{table:2}, and for $k=5$ and $\lambda=5$, point E exists and is an unstable non-hyperbolic critical point which is obvious by its vector plot and in accordance with Figure~\ref{fig:E1F1G1}.  Figure~\ref{fig:E1F1G1} illustrates the second scaling solution for $k=4$ and $\lambda=3$. For that particular set of the parameters point F is hyperbolic and behaves as a saddle critical point. Point F is defined by the red dot with coordinates $x_1\simeq0.4,x_2\simeq0.4,x_3=0$ .

Figure~\ref{fig:E1F1G1} demonstrates the three-dimensional phase space of $x_1,x_2,x_3$ when $k=2$ and $\lambda=1$. Blue arrows denote the vector field while the critical point is defined by a red dot with coordinates $x_1\simeq0.41,x_2\simeq0.91,x_3=0$.  As stated in Tables~\ref{table:1} and \ref{table:2}, and for that specific set of the parameters, point G exists and behaves as a saddle hyperbolic critical point in accordance with Figure~\ref{fig:imG}.

\begin{acknowledgments}
The work was supported by Nazarbayev University Faculty Development Competitive Research Grant No. 11022021FD2926 and by the Hellenic Foundation for Research and Innovation (H.F.R.I.) under the “First Call for H.F.R.I. Research Projects to support Faculty members and Researchers and the procurement of high-cost research equipment grant” (Project Number: 2251). This article is based upon work from COST Action CA21136 Addressing observational tensions in cosmology with systematics and fundamental physics (CosmoVerse) supported by COST (European Cooperation in Science and Technology).
\end{acknowledgments}

\bibliographystyle{utphys}
\bibliography{references}

\end{document}